\documentclass[conference]{IEEEtran}
\IEEEoverridecommandlockouts
\usepackage{cite}
\usepackage[hidelinks]{hyperref}
\usepackage{amsmath,amssymb,amsfonts}
\usepackage{xcolor} % Load the color package
\usepackage{algorithm}
\usepackage{algorithmic}
\usepackage{graphicx}
\usepackage{subfig}
\usepackage{multicol}
\usepackage{multirow}
\usepackage{array} % For centering text in tables
\usepackage{colortbl} % For coloring table cells
\usepackage[normalem]{ulem}
\usepackage{enumitem}
\usepackage{url}
\usepackage{setspace}

\def\BibTeX{{\rm B\kern-.05em{\sc i\kern-.025em b}\kern-.08em
    T\kern-.1667em\lower.7ex\hbox{E}\kern-.125emX}}

\definecolor{lightgreen}{RGB}{144,238,144}
\definecolor{lightyellow}{RGB}{255,255,224}
\definecolor{lightred}{RGB}{255,182,193}
\definecolor{custompurple}{HTML}{e05070}

\newcommand{\bluehighlight}[1]{\colorbox[HTML]{dae8fc}{#1}} % Highlighting for the sections 
\newcommand{\lightredhighlight}[1]{\colorbox[HTML]{ffe6e6}{#1}} % Light red highlight
\newcommand{\deeperredhighlight}[1]{\colorbox[HTML]{ff8080}{#1}} % Deeper red highlight
\newcommand{\darkestredhighlight}[1]{\colorbox[HTML]{ff1a1a}{\textcolor{black}{#1}}} % Darkest red highlight
\newcommand{\greenhighlight}[1]{\colorbox[HTML]{1aff1a}{#1}} % Green highlight
\newcommand{\yellowhighlight}[1]{\colorbox[HTML]{fff49e}{#1}} % Yellow highlight

\newcommand{\algName}{\textbf{Cute-Lock}}
\newcommand{\behave}{\textbf{\algName-Beh}}
\newcommand{\struct}{\textbf{\algName-Str}}

\begin{document}

\title{Cute-Lock: Behavioral and Structural Multi-Key Logic Locking Using Time Base Keys}

\author{\IEEEauthorblockN{Kevin Lopez}
\IEEEauthorblockA{\textit{Computer Engineering \& Computer Science Department} \\
\textit{California State University, Long Beach}\\
Long Beach, CA, USA \\
Kevin.LopezChavez01@student.csulb.edu}
\and
\IEEEauthorblockN{Amin Rezaei}
\IEEEauthorblockA{\textit{Computer Engineering \& Computer Science Department} \\
\textit{California State University Long Beach}\\
Long Beach, CA, USA \\
Amin.rezaei@csulb.edu} 
}

\maketitle

\begin{abstract}
The outsourcing of semiconductor manufacturing raises security risks, such as piracy and overproduction of hardware intellectual property. To overcome this challenge, logic locking has emerged to lock a given circuit using additional key bits. While single-key logic locking approaches have demonstrated serious vulnerability to a wide range of attacks, multi-key solutions, if carefully designed, can provide a reliable defense against not only oracle-guided logic attacks, but also removal and dataflow attacks. In this paper, using time base keys, we propose, implement and evaluate a family of secure multi-key logic locking algorithms called \algName~that can be applied both in RTL-level behavioral and netlist-level structural representations of sequential circuits. Our extensive experimental results under a diverse range of attacks confirm that, compared to vulnerable state-of-the-art methods, employing the \algName~family drives attacking attempts to a dead end without additional overhead. \\
\end{abstract}


\section{Introduction} 
\label{Sec:Intro}
As chip manufacturing becomes increasingly outsourced, the need for robust protection mechanisms is constantly increasing. In this split of design and manufacturing, one company designs the digital Integrated Circuit (IC), while another handles the physical fabrication phase. Logic locking \cite{EndingPiracy, 1fault_analysis, HARPOON} has emerged as a promising solution to prevent piracy and overproduction of ICs. Combinational locking is the process of adding additional inputs, called key bits, to an IC to corrupt the output when the incorrect key is inserted. On the other hand, sequential locking is the procedure of adding additional obfuscation states where the user needs to traverse through those states to utilize the circuit.

Traditional logic locking solutions are susceptible to the oracle-guided SAT attack \cite{evaluatingLogicEncryptionAlgorithms} that extracts the key using an oracle (i.e., a working chip bought off the market) and a locked netlist, potentially leaked from an untrustworthy foundry. Although different logic locking solutions have been proposed to protect against the original SAT attack \cite{AntiSAT, SARLock}, they are still susceptible to approximate versions of oracle-guided SAT attacks \cite{AppSat, DoubleDIP} as well as dataflow and removal attacks \cite{dana, fall} that try to remove or reverse engineer the added lock and extract the original circuit.

While the majority of the logic locking solutions have been focused on single-key methods, we believe that their vulnerabilities against powerful attacks can be effectively mitigated through time-based multi-key logic locking. Thus, in this paper, we introduce the idea of advanced multi-key logic locking for sequential circuits in which the states are locked using a counter and multiple keys that are applied at different times. For the circuit to operate correctly, these key values must be provided in a specific order determined by the corresponding clock cycle. The motivation is to improve security by complicating the decoding process, ensuring that the circuit only operates as intended when the correct key sequence is applied at the correct time. Our proposed approach uses the current hardware to manipulate the state transition and provides a low-overhead solution for sequential logic locking. With the addition of multiple keys, it maintains resilience against oracle-guided attacks with \cite{evaluatingLogicEncryptionAlgorithms, DoubleDIP, AppSat} or without \cite{BMC, KC2, RANE} scan chain access assumption. Additionally, it improves structural integrity against dataflow \cite{dana} and removal \cite{fall} attacks. In this paper, we present the following contributions: 
\begin{enumerate}
    \item [$\bullet$] Proposing \behave, a \textbf{\uline{c}}ounter-based m\textbf{\uline{u}}l\textbf{\uline{t}}i-k\textbf{\uline{e}}y logic \textbf{\uline{lock}}ing \textbf{\uline{beh}}avioral solution to secure circuits against logic attacks in RTL-level representation;
    \item [$\bullet$] Proposing \struct, a \textbf{\uline{c}}ounter-based m\textbf{\uline{u}}l\textbf{\uline{t}}i-k\textbf{\uline{e}}y logic \textbf{\uline{lock}}ing  \textbf{\uline{str}}uctural solution to secure circuits against both logic and structural attacks in the gate-level netlist;
    \item [$\bullet$] Generating more than 60 benchmarks based on the proposed methods, evaluating their overhead, and testing their security against state-of-the-art oracle-guided as well as removal and dataflow attacks.
\end{enumerate}

%The rest of the paper is organized as follows: Section \ref{Sec:Related} discusses the background and related works, including existing efforts on multi-key logic locking. Section \ref{Sec:Contribution} proposes our behavioral and structural multi-key logic locking solutions. Section \ref{Sec:Experiments} depicts the extensive experimental results on the security and overhead of the proposed methods. Finally, conclusions are given in Section \ref{Sec:Conclusion}. 

\section{Background and Related Work} 
\label{Sec:Related}
In this section, we begin by examining the development of logic locking methods alongside oracle-guided and removal attacks, followed by an overview of current approaches in multi-key logic locking.

\subsection{Logic Locking Techniques} 
\label{Sec:Defenses}
Logic locking techniques can be categorized into combinational locking and sequential locking. Initial combinational logic locking solutions were {\sc xor}-based and {\sc mux}-based mechanisms \cite{EndingPiracy, 1fault_analysis}. However, the oracle-guided SAT attack \cite{evaluatingLogicEncryptionAlgorithms} has exposed vulnerabilities in these methods, leading to the development of more robust techniques such as Anti-SAT \cite{AntiSAT}, SAR-Lock \cite{SARLock}, TT-Lock \cite{TTlock}, SFLL \cite{SFLL}, BLE \cite{rezaei2020rescuing}, DLE \cite{6_Distributed_logic_encryption}, and others \cite{7_Global_attack, 9_Sequential_logic_encryption, 10_CoLA, 5_CyclicL} that increase the time complexity of SAT attacks. Furthermore, HARPOON \cite{HARPOON} is a sequential logic locking solution that adds additional states to provide an obfuscation mode; in order to utilize the circuit, one must navigate through the obfuscation mode first. Boosted Finite-State Machine (BFSM) \cite{BSFM} uses a Physical Unclonable Function (PUF) to determine the starting state, which is typically an obfuscated state; to unlock and use the circuit, a specific input sequence must be provided to traverse from the obfuscated state to the functional state. In addition, Dynamic State Reflection (DSR) \cite{DSD} adds black holes to hide the boundary between obfuscation and functional modes. While all these logic locking methods have been challenged by different attacks such as SAT attack \cite{evaluatingLogicEncryptionAlgorithms}, Key-Condition Crunching (KC2) \cite{KC2}, Bounded Model Checking (BMC) \cite{BMC}, and Reverse Assessment of
Netlist Encryption (RANE) \cite{RANE}, researchers have explored other combinational and sequential methods \cite{2__FULL_LOCK, karmakar2019efficient, HLock, zhou2019resolving, Fortifying_RTL_locking} as well.

Adding extra states would incur additional overhead when using the circuit since one would need to transition between the obfuscated states. In addition, single-key solutions remain susceptible once the key is compromised, potentially exposing the entire security of the IC. In our proposed solutions, different keys need to be provided to the circuit at different times, and they re-route users to the wrong states whenever a wrong key is applied. In this case, the circuit starts at the initial state and will maintain the same state transition as the original circuit as long as the correct keys are applied.

\subsection{Logic Locking Attacks} 
Logic locking attacks are categorized into oracle-guided logic attacks as well as removal attacks. The oracle-guided SAT attack \cite{evaluatingLogicEncryptionAlgorithms} is one of the most powerful attacks against combinational locking and sequential locking with scan access; it uses a SAT solver to iteratively find Discriminating Input Patterns (DIPs) that eliminate incorrect keys, eventually converging on the correct key. This attack has been shown to break almost all of the early logic locking schemes efficiently \cite{EndingPiracy, 1fault_analysis, HARPOON}. An extension of the SAT attack, AppSAT \cite{AppSat} aims to find approximate keys that work for most input patterns. This attack is particularly effective against schemes such as Anti-SAT \cite{AntiSAT} with low output corruptibility. The Double DIP attack \cite{DoubleDIP} improves upon the SAT attack by finding two DIPs in each iteration, allowing it to break certain SAT-resistant locking schemes such as SAR-Lock \cite{SARLock}.

On the sequential side, the BMC attack \cite{BMC} targets sequential logic locking by unrolling the circuit for a fixed number of time steps and using a SAT solver to find the key. The KC2 attack \cite{KC2} improves upon the BMC attack by using incremental SAT solving and dynamic simplification of key conditions. The RANE attack \cite{RANE} uses API-based invocation of formal verification tools to model the initial state as a secret key variable and find the unlocking sequence. 

In addition, the Functional Analysis Attack on Logic Locking (FALL) \cite{fall} is primarily a removal attack, with some structural components designed to extract the logic locking key. FALL has shown to be successful against logic locking techniques such as TTLock \cite{TTlock} and SFLL \cite{SFLL}. In addition, the Dataflow-based Netlist Analysis (DANA) attack \cite{dana} is designed to assist reverse engineering circuits by structuring an unstructured sea of gates. The key operation is to group registers into distinct clusters, which can then be analyzed to derive the high-level architecture and functionality of the circuit. By focusing on the flow of data between Flip-Flops (FFs), DANA helps to recover meaningful high-level structures from a flatten netlist, making it a crucial first step in the reverse engineering process. Other attacks have also been proposed \cite{3__ATTACK_ExposesYou, 3__ATTACK_ORACALL, Rezaei:BreakUnroll, Rezaei:LIPSTICK} targeting a subset of existing logic locking solutions.

\subsection{Multi-Key Approaches}
\label{Sec:MultiKey}
Recently, the idea of using multi-key solutions to mitigate oracle-guided SAT attacks is proposed \cite{DKLock, 10808395, SLED, GateLock, Rezaei:KGate}. DK-lock \cite{DKLock} is a two-key logic locking solution that uses an activation key and then a functional key. The control logics of both the activation phase and the functional phase then unified into an FSM \cite{10808395}. DK-Lock is susceptible to attacks like \cite{Rezaei:BreakUnroll} that can expand the key size to reverse the method back to a single-key solution. SLED \cite{SLED} is another multi-key sequential solution that works by dynamically changing the keys during circuit operation generated by a secure module based on a static seed. However, since it depends on a seed value to operate, it is vulnerable to oracle-guided SAT attacks if the attacker deciphers the seed value as the initial key. Gate-Lock \cite{GateLock} uses an approach focused on locking gates; the resulting key to use the circuit changes depending on the input needed. In addition, K-Gate Lock \cite{Rezaei:KGate} is based on input encoding and can be fully implemented using combinational logic without the need for state-holder components. However, neither provides any structural benefit against dataflow and removal attacks. 

While each state-of-the-art multi-key logic locking method aims to secure against a different attack, our goal in this paper is to propose low-overhead multi-key solutions at both the RTL-level and netlist-level that are secure against all the above-mentioned attacks.

\section{Cute-Lock Family} 
\label{Sec:Contribution}
In this section, after explaining the terminology, we discuss our proposed methods of locking a circuit with multiple keys; the first one, called \behave, is an RTL-level sequential logic locking approach that is secure against oracle-guided attacks and requires different key values based on different clock cycles to operate correctly. The second method, called \struct, is a netlist-level implementation of our behavioral solution that resists not only oracle-guided SAT attacks but also dataflow and removal attacks.

\subsection{Terminology}
\label{Sec:Terminology}
In the context of \algName~family, it is crucial to understand the terminology used to describe the components:
\begin{description}
    \item[n:] Number of inputs to the circuit.
    \item[k:] Number of key values (multiple keys, in contrast to traditional logic locking algorithms that used a single key). 
    \item[k$_i$:] Number of bits in each key value (i.e., the key size). %For \struct, k$_i$ is determined by the number of select lines ($\|select\|$) in the first multiplexer.
    \item[c:] Number of clock cycles for the counter, determining when specific keys must be provided. % 
    \item[m:] Number of layers in the {\sc mux} tree equal to $\log_2(k) + 1$.
    \end{description}

\subsection{Behavioral Solution} 
\label{Sec:Behavioral count lock}
\begin{figure}[!t]
    \centering
    \includegraphics[width=\columnwidth]{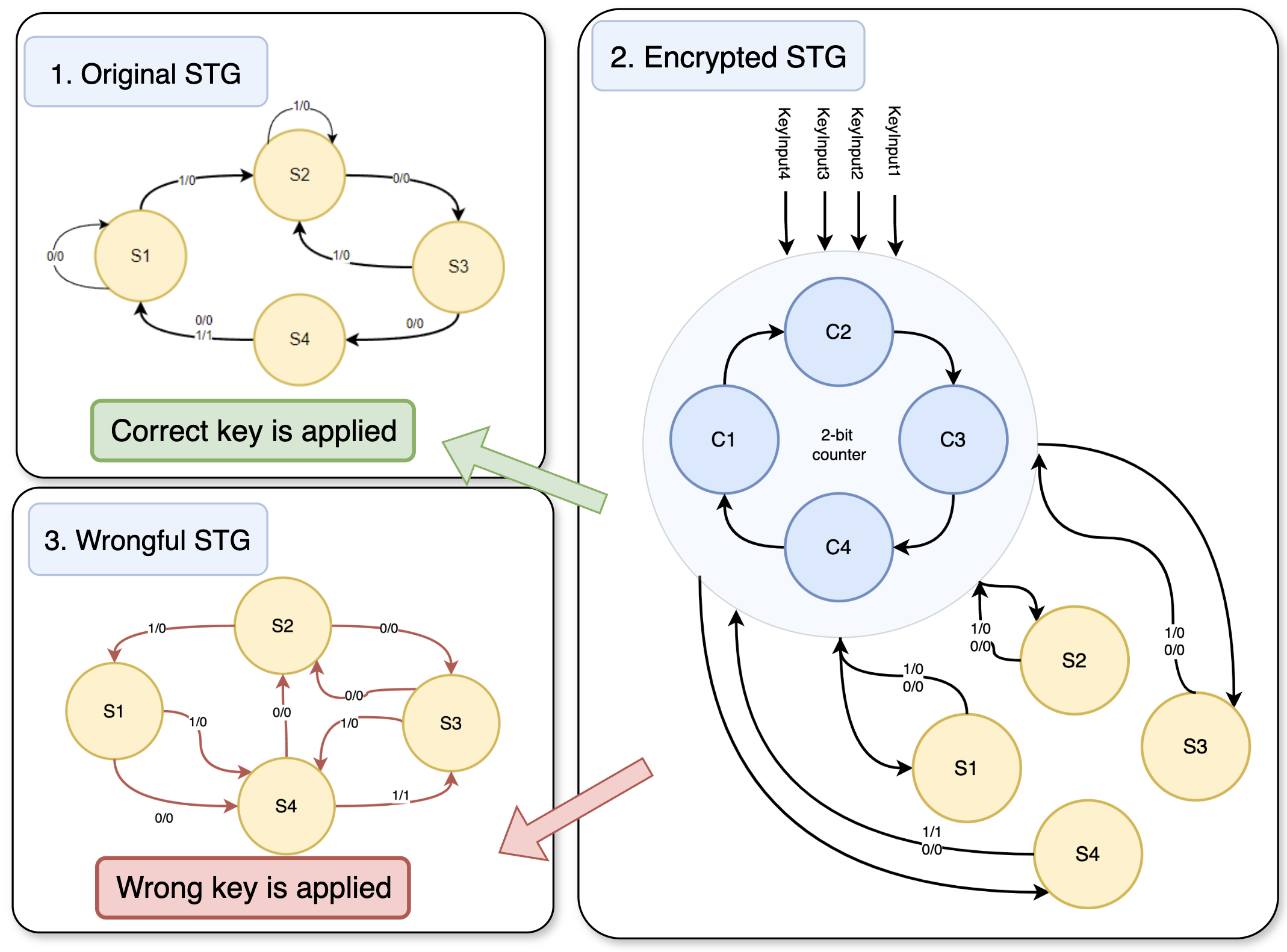}
    \caption{\behave~STG example}
    \label{fig: behave stg}
\end{figure}

For RTL-level sequential circuit, the core idea is to require a specific key value based on the time count for the circuit to behave as intended. When an incorrect key is provided to a particular clock cycle, the circuit transitions to a random, incorrect state. Fig. \ref{fig: behave stg} demonstrates how \behave~affects the high-level State Transition Graph (STG) using a 1001 sequence detector example. Here, we can see that the keys and counter work together to handle the state transition for the next state, and whenever the wrong key is provided, a wrongful state transition is taken. Implementing \behave~requires minimal changes to the RTL code. The only additions are a counter and the wrongful state transitions when a wrong key is provided. These wrongful state transitions are added to the FF logic, where the present state updates occur. Here are the components of the sequence detector example in Fig. \ref{fig: behave stg}:

\begin{description}
    \item[\bluehighlight{1. Original STG:}]The original STG for detecting a 1001 sequence as a mealy machine. Also, it serves as the same STG whenever the correct keys are provided.
    \item[\bluehighlight{2. Encrypted STG:}]The encrypted STG with four keys, 4 bits each, and a 2-bit counter. The correct state transition would be determined by the counter, along with providing the proper key. If the wrong key is provided, a wrongful state transition is taken. 
    \item[\bluehighlight{3. Wrongful STG:}]The incorrect STG is constructed of random state transitions defined at the RTL-level. 
\end{description}

\begin{figure}[!b]
    \centering
    \includegraphics[width=\columnwidth]{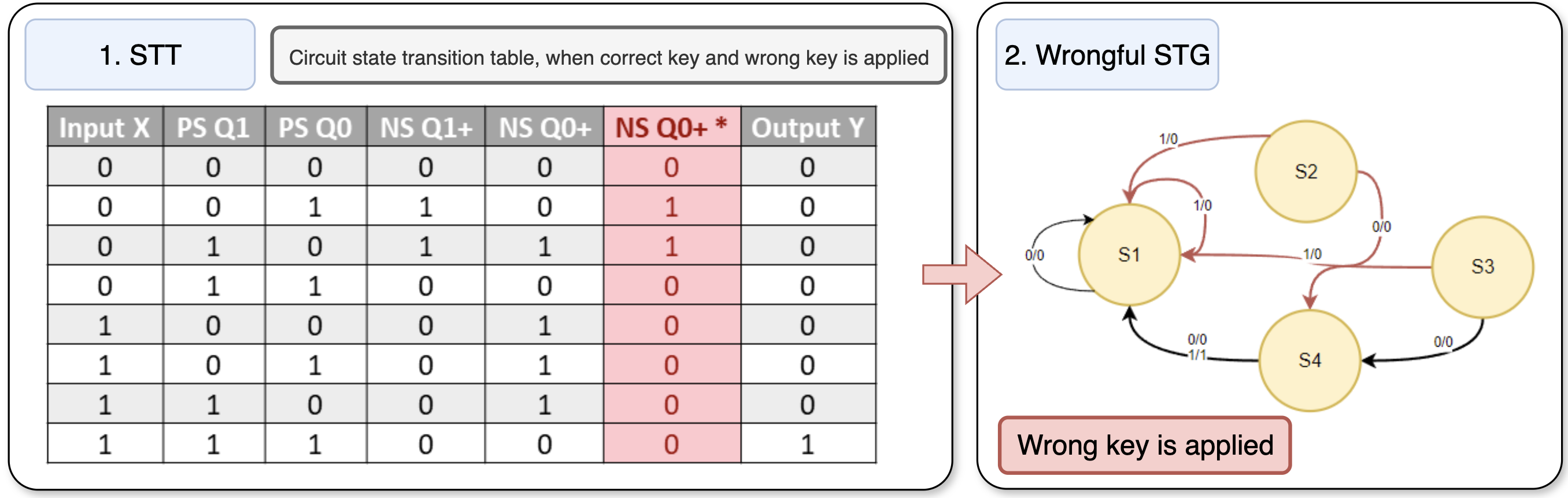}
    \caption{\struct~STG \& STT example}
    \label{fig: struct state transtion}
\end{figure}

While \behave~does not represent a significant change in the RTL-level representation, if one plans to convert a given netlist back to STG format, they face the state explosion problem \cite{Valmari1998}. Thus, currently, we have used Xilinx Vivado to implement \behave~that use {\sc mux}s instead of redesigning the STG from the ground up. However, in this case, while it provides security against oracle-guided SAT attacks, it does not provide substantial structural benefits against removal attacks. Thus, it requires another step of obfuscation on top of the behavioral method. To address this issue, we discuss a more efficient structural solution with lower overhead in Section \ref{Sec:structural count lock}.

\subsection{Structural Solution} 
\label{Sec:structural count lock}
In \struct, instead of transitioning to a random state upon wrong key insertion, it moves to a different state predefined by existing state transitions. Fig.~\ref{fig: struct state transtion} demonstrates how \struct~affects the State Transition Table (STT) using the same 1001 sequence detector used in Fig. \ref{fig: behave stg}:
\begin{description}
    \item[\bluehighlight{1. STT:}] The STT for the 1001 sequencer detector along with wrongful state transition. The highlighted column (NS Q0+*) indicates values for ``NS Q0+'' whenever the wrong keys are provided. This will make up the new wrongful state transition. In this case, the hardware from ``NS Q1+'' is repurposed to be used on wrongful state transition. For example, if the current state (PS Q1, PS Q0) is ``00'' and input $X$ is 1, the next state (NS Q1+, NS Q0+) should be ``01'', but when a wrong key-counter combination is given, it will stay at state ``00''.
    \item[\bluehighlight{2. Wrongful STG:}] The incorrect STG is constructed from ``NS Q1+'' and ``NS Q0+*''. 
\end{description}

\begin{figure}[!t]
   \centering
    \includegraphics[width=1\linewidth]{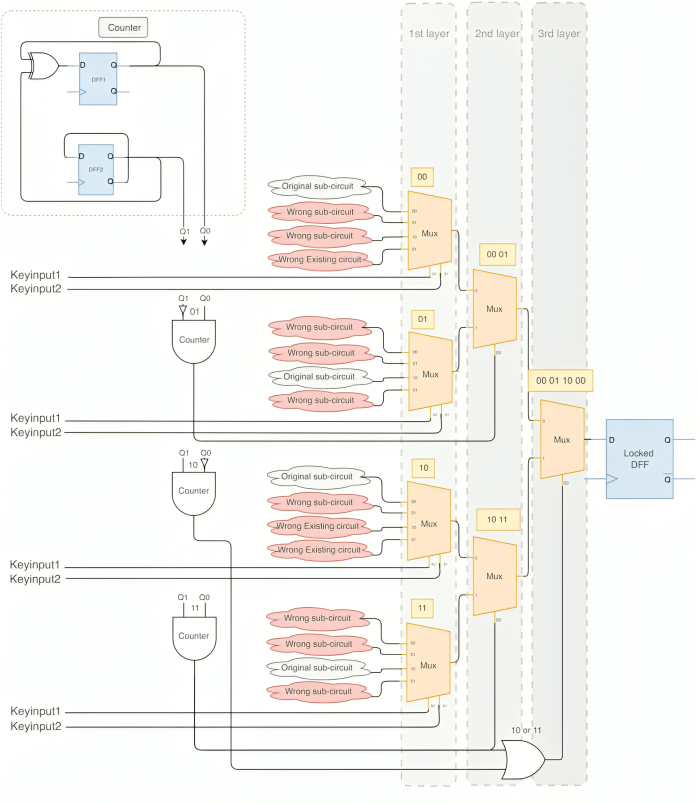}
    \caption{\struct~{\sc mux} tree example  }
    \label{fig: Multiplexer Tree}
\end{figure}

The FFs in the structural solution are connected through a tree of {\sc mux}s illustrated in Fig.~\ref{fig: Multiplexer Tree}. The gray clouds show the correct hardware for the FF to perform the state transition correctly, and the red clouds show the hardware of another FF that would accomplish the wrongful state transition. The {\sc mux} tree allows the ability to lock an FF with multiple keys synchronized with a counter and enables the FFs to utilize incorrect hardware when the wrong key is provided at a specific time. The {\sc mux} tree is made up of $m$ layers, where $m$ is $log_2(k)+1$. Each layer consists of the following: 
\begin{description}
    \item[$\mathbf{1st}$ Layer:] The first layer of {\sc mux}s verifies that the correct key has been provided, denoted by $keyinput_1$ to $keyinput_{k_i}$. This layer also defines the key size $k_i$, which is determined by the $select$ size of the multiplexer. The number of different wrongful hardware configurations is given by $2^{k_i}-1$. In Fig.~\ref{fig: Multiplexer Tree}, the keys are 00, 10, 00, and 10 in respective counter times. The key sizes and wrongful hardware are denoted by the 4-to-1 {\sc mux}s. 
    \item[$\mathbf{2nd, 3rd, ..., (m-1)th}$ Layers] These layers are controlled by the counter to select the correct {\sc mux}. The counter values are divided at each subsequent layer. The {\sc mux}s at these layers determine whether to use the top or bottom connection based on the previous counter values applied in the bottom {\sc mux}s. The check is performed by {\sc or}-ing all the counter times in the previous {\sc mux}s and feeding the result into the $select$ input of the current {\sc mux}.
     \item[$\mathbf{mth}$ Layer] The output of the final {\sc mux} is fed into the FF.
\end{description}

The {\sc mux}-tree has a width of $log_2(k)+1$ and a height of $k$ where $k$ is the number of distinct keys to lock the FFs. When using \struct, it is possible to lock any number of FFs. While locking one FF with different keys is enough to resist oracle-guided SAT attacks, locking more FFs would provide more resilience against dataflow and removal attacks.

\section{Experimental Results} 
\label{Sec:Experiments}
We conduct experiments on a Windows 11 machine, which accesses Linux Ubuntu 22.04 via WSL2. The machine is an Intel 13900H with 14 cores and 32 threads at 2.6 GHz and 56 GB of DDR5 RAM. First, we validate the \algName~family and then test them against oracle-guided SAT attacks. Next we test \struct~against dataflow and removal attacks and finally implement it on Cadence Genus to analyze and compare the overhead with state-of-the-art works. 
% ---------------------- Manually move this to the bottom of the page
\begin{minipage}[!b]{\linewidth} %
    \vspace{10pt}
    \noindent\rule[0.5ex]{0.75\linewidth}{0.4pt} % Horizontal line, 75% of text width
    
    \noindent
    \footnotesize
    % these footnotes appear horizontally ( 1, 2, 3, 4, 5) 
    \footnotemark[1] \lightredhighlight{\phantom{x}} \cellcolor{red!10}CNS, \hspace{2mm}
    \footnotemark[2] \deeperredhighlight{\phantom{x}} \cellcolor{red!50}x..x,   \hspace{2mm}
    \footnotemark[3] \darkestredhighlight{\phantom{x}} \cellcolor{red!90}FAIL,   \hspace{2mm}
    \footnotemark[4] \greenhighlight{\phantom{x}} \cellcolor{green!100}Equal  \hspace{2mm}
    \footnotemark[5] \yellowhighlight{\phantom{x}} \cellcolor{yellow!100}N/A\hspace{2mm}
    \end{minipage}
% ---------------------- to here

In the tables, different colors are used to indicate specific conditions. The color light red\footnotemark[1] represents the ``condition not solvable'' status. A deeper red\footnotemark[2] signifies a wrong key, while the darkest red\footnotemark[3] indicates that the attack failed. Green\footnotemark[4] denotes that the correct key has been found. Yellow\footnotemark[5] means the attack did not report any key within the time limit of 20 hours. The source codes and created benchmarks of \algName~family are publicly available on our GitHub repository\footnote{https://github.com/cars-lab-repo/Cute-Lock}.

\subsection{Algorithm Validation}
The validation of \algName~is done in Xilinx Vivado. Locking benchmarks with the same key values (i.e., reduced to a single-key solution) leads to SAT attacks from NEOS~\cite{neos} and RANE~\cite{RANE} to find the correct key as expected.

 \begin{table}[!h]
        \caption{\behave~validation}
        \centering
        \scriptsize
            \begin{tabular}{|c|c|c|c|c|c|}
            \hline
            \textbf{ } & \multicolumn{1}{|c|}{\textbf{Inputs}} & \multicolumn{3}{|c|}{\textbf{ Outputs }} \\
            \hline
            \textbf{Time (ns)} & \textbf{x[7:0]} & \textbf{y[38:0]} & \textbf{$y_{ck}[38:0]$} & \textbf{$y_{wk}[38:0]$}  \\
            \hline
            0 & 0 &   0 & \cellcolor{green!50} 0 & 0 \\
            60 & 2aaaa &  0 & \cellcolor{green!50} 0 &  400000000 \\
            100 &  3c3c3  &  2000002007  & \cellcolor{green!50} 2000002007  & 00000000e \\
            120 & 3c3c3  &   1800000002 & \cellcolor{green!50} 1800000002 &  00000000e \\
            160 & 2aaaa  &  0 & \cellcolor{green!50} 0 & 00000000e \\
            200 & 3c3c3  &  1800000002 & \cellcolor{green!50} 1800000002 &  000071 \\
            220 & 2aaaa  &  400240 & \cellcolor{green!50} 400240 &  91 \\
            240 & 2aaaa  &  0 & \cellcolor{green!50} 0 &  91 \\
            260 & 2aaaa  &  0 & \cellcolor{green!50} 0 &  91 \\
            280 & 2aaaa  &  0 & \cellcolor{green!50} 0 &  2004 \\
            300 & 0  &  0 & \cellcolor{green!50} 0 & 2004 \\
            320 & 0  &  0 & \cellcolor{green!50} 0 & 0 \\
            340 & 0  &  0 & \cellcolor{green!50} 0 & 0 \\
            360 & 0  &  0 & \cellcolor{green!50} 0 &  0 \\
            380 & 0  &  20000002007 & \cellcolor{green!50} 20000002007 &  e \\
            380 & 3c3c3  &  1800000002 & \cellcolor{green!50} 1800000002 & e \\
            \hline
        \end{tabular}
        \label{tab: behave validation}
    \end{table}
    
\subsubsection{Behavioral Solution}
For \behave~algorithm validation, we lock the \textit{bcomp} benchmark from the Synthezza suite~\cite{FSMs} with 19 key-bit values. When the correct key values are provided, both the original and the locked circuit behave the same, as shown in Table~\ref{tab: behave validation}, where columns $y_{ck}$ and $y_{wk}$ are the outputs under the correct and wrong keys, respectively.

 \begin{table}[!h]
        \caption{\struct~validation}
        \centering
        \scriptsize
        \begin{tabular}{|c|c|c|c|c|c|c|c|}
        \hline
        \textbf{ } & \multicolumn{4}{|c|}{\textbf{Inputs}} & \multicolumn{3}{|c|}{\textbf{ Outputs }} \\
        \hline
        \textbf{Time (ns)} & \textbf{G0} & \textbf{G1} & \textbf{G2} & \textbf{G3} & \textbf{G17} & \textbf{$G17_{ck}$} & \textbf{$G17_{wk}$} \\
        \hline
        0   & 0 & 1 & 0 & 1 & x & \cellcolor{green!50}  x &  x  \\
        20  & 1 & 0 & 1 & 0 & 1 & \cellcolor{green!50}  1 & 1  \\
        40  & 1 & 1 & 0 & 0 & 1 & \cellcolor{green!50}  1 &  1  \\
        60  & 1 & 1 & 1 & 0 & 1 & \cellcolor{green!50}  1 & 1  \\
        80  & 0 & 1 & 0 & 1 & 1 & \cellcolor{green!50}  1 & 1  \\
        100 & 1 & 0 & 1 & 0 & 1 & \cellcolor{green!50}  1 &  1  \\
        120 & 0 & 0 & 0 & 0 & 1 & \cellcolor{green!50}  1 &  1  \\
        140 & 1 & 1 & 1 & 1 & 1 & \cellcolor{green!50}  1 &  0  \\
        160 & 0 & 0 & 1 & 1 & 0 & \cellcolor{green!50}  0 & 0  \\
        180 & 1 & 0 & 0 & 1 & 0 & \cellcolor{green!50}  0 &  1  \\
        200 & 0 & 1 & 1 & 0 & 0 & \cellcolor{green!50}  0 &  0  \\
        220 & 0 & 1 & 1 & 1 & 0 & \cellcolor{green!50}  0 & 1  \\
        240 & 1 & 1 & 0 & 1 & 1 & \cellcolor{green!50}  1 &  1  \\
        260 & 0 & 0 & 0 & 1 & 1 & \cellcolor{green!50}  1 &  1  \\
        280 & 1 & 0 & 1 & 1 & 1 & \cellcolor{green!50}  1 &  0  \\
         \hline
        \end{tabular}
        \label{tab:structural_algorithm_validation}
        \end{table}

\subsubsection{Structural Solution}
For the \struct~algorithm validation, \textit{s27} from ISCAS’89 \cite{iscas85} is locked using the following keys: 1, 3, 2, 0. When the correct key values are provided, the original and the locked circuit behave equally, as shown in Table~\ref{tab:structural_algorithm_validation}, where $G17_{ck}$ and $G17_{wk}$ represent the output of the circuit when the correct keys and the wrong keys are provided, respectively.

\subsection{Logic Attacks Evaluation}
\label{Sec:Logic-Attacks-Eval}
One of the main objectives of \algName~family is to generate a locking mechanism that oracle-guided SAT attacks will not be able to decrypt. In this section, we will test \behave~and \struct~against SAT-based oracle-guided attacks.  

\subsubsection{Behavioral Solution} 
To test \behave, we generate locked versions of the Synthezza benchmark suite~\cite{FSMs} in Verilog format, then use Yosys~\cite{yosys} to convert to .blif format. While the files are in .blif format, it is necessary to convert some FFs into latches. Then, ABC~\cite{abc} is used to convert to .bench format, which is used to run the SAT attacks using NEOS~\cite{neos}. None of the benchmarks run provide the correct keys as shown in Table~\ref{tab: behave against oracle}.
        
        \begin{table}[!t]
            \caption{\behave~security against logic attacks}
            \centering
            \setlength{\tabcolsep}{3pt} % Reduce the space between columns
            \tiny
             \begin{tabular}{|c|l|c|c|c|c|c|}
                \hline
                 \multicolumn{4}{|c|}{\textbf{Benchmark and Locking Information}} & \multicolumn{3}{|c|}{\textbf{NEOS~\cite{neos}}} \\
                \hline
                 \textbf{Synthezza~\cite{FSMs}} & \textbf{Circuit} & \textbf{\# Keys ($k$) } & \tiny{\textbf{Key Size ($k_i$)}} & \textbf{BBO} & \textbf{INT} & \textbf{KC2} \\
                \hline
                \multirow{12}{*}{\rotatebox[origin=c]{90}{\textbf{Small}}}
                  &  bcomp & 6 & 18 & \cellcolor{red!50} 6m25.446s & \cellcolor{red!50} 0m0.885s & \cellcolor{red!50} 0m1.030s \\  
                  &  bech & 6 & 18 & \cellcolor{red!50} 6m4.845s & \cellcolor{red!50} 0m0.723s & \cellcolor{red!50} 0m0.838s \\  
                  &  bridge & 5 & 16 & \cellcolor{red!50} 3m28.614s & \cellcolor{red!90} 0m0.100s & \cellcolor{red!90} 0m0.182s \\  
                  &  cat & 3 & 11 & \cellcolor{red!50} 15m1.161s & \cellcolor{red!50} 0m0.772s & \cellcolor{red!50} 0m0.680s \\  
                  &  checker9 & 3 & 10 & \cellcolor{red!50} 3m0.931s & \cellcolor{red!50} 0m0.842s & \cellcolor{red!50} 0m0.803s \\  
                  &  cpu & 4 & 14 & \cellcolor{red!50} 2m11.658s & \cellcolor{red!50} 0m0.732s & \cellcolor{red!50} 0m0.799s \\  
                  &  dmac & 2 & 7 & \cellcolor{red!50} 1m45.751s & \cellcolor{red!50} 0m0.623s & \cellcolor{red!50} 0m0.681s \\  
                  &  e10 & 3 & 10 & \cellcolor{red!50} 3m17.832s & \cellcolor{red!50} 0m0.816s & \cellcolor{red!50} 0m1.033s \\  
                  &  e15 & 4 & 13 & \cellcolor{red!50} 8m59.511s & \cellcolor{red!50} 0m1.361s & \cellcolor{red!50} 0m1.462s \\  
                  &  e16 & 4 & 13 & \cellcolor{red!50} 7m50.966s & \cellcolor{red!50} 0m0.774s & \cellcolor{red!50} 0m0.918s \\  
                  &  e161 & 5 & 16 & \cellcolor{red!50} 2m53.761s & \cellcolor{red!50} 0m0.731s & \cellcolor{red!50} 0m0.759s \\  
                  &  e17 & 2 & 8 & \cellcolor{red!50} 15m0.543s & \cellcolor{red!50} 0m0.522s & \cellcolor{red!50}0m0.607s \\ \hline  
                \multirow{15}{*}{\rotatebox[origin=c]{90}{\textbf{ Medium}}}  
                  & acdl & 5 & 16 & \cellcolor{red!50}14m47.149s & \cellcolor{red!90} 0m0.641s & \cellcolor{red!50} 0m1.157s \\  
                 & alf & 0 & 31 & \cellcolor{red!90}0m0.180s & \cellcolor{red!90} 0m0.107s & \cellcolor{red!50} 0m0.469s \\  
                 & amtz & 7 & 23 & \cellcolor{red!50}14m9.747s & \cellcolor{red!50} 0m2.227s & \cellcolor{red!50} 0m2.727s \\  
                 & ball & 4 & 44 & \cellcolor{red!50}15m5.744s & \cellcolor{red!90} 0m1.162s & \cellcolor{red!90} 0m4.998s \\  
                 & bens & 7 & 21 & \cellcolor{red!50}15m3.290s & \cellcolor{red!50} 0m18.365s & \cellcolor{red!50} 0m19.804s \\  
                 & berg & 7 & 21 & \cellcolor{red!50}10m5.736s & \cellcolor{red!50} 0m1.730s & \cellcolor{red!50} 0m2.531s \\  
                 & bib & 7 & 21 & \cellcolor{red!50}15m3.795s & \cellcolor{red!50} 0m2.750s & \cellcolor{red!50} 0m3.324s \\  
                 & big & 6 & 18 & \cellcolor{red!50}11m14.492s & \cellcolor{red!90} 0m0.658s & \cellcolor{red!90} 0m1.163s \\  
                 & bs & 6 & 19 & \cellcolor{red!50}9m52.679s & \cellcolor{red!90} 0m0.600s & \cellcolor{red!90} 0m0.895s \\  
                 & codec & 2 & 4 & \cellcolor{red!50}15m2.282s & \cellcolor{red!50} 0m2.252s & \cellcolor{red!50} 0m2.032s \\  
                 & codec1\_2 & 9 & 28 & \cellcolor{red!50}15m4.756s & \cellcolor{red!50} 0m3.768s & \cellcolor{red!50} 0m4.005s \\  
                 & cow & 6 & 49 & \cellcolor{red!50}15m7.930s & \cellcolor{red!90} 0m1.225s & \cellcolor{red!90} 0m4.587s \\  
                 & cyr & 6 & 20 & \cellcolor{red!50}14m7.341s & \cellcolor{red!50} 0m2.375s & \cellcolor{red!50} 0m3.072s \\  
                 & dav & 6 & 18 & \cellcolor{red!50}15m3.939s & \cellcolor{red!90} 0m0.519s & \cellcolor{red!90} 0m1.019s \\  
                 & doron & 7 & 22 & \cellcolor{red!50}13m49.117s & \cellcolor{red!50} 0m2.854s &\cellcolor{red!50} 0m4.004s \\  \hline
                \multirow{6}{*}{\rotatebox[origin=c]{90}{\textbf{Large}}}  
                 & absurd & 21 & 65 & \cellcolor{red!50} 15m25.370s & \cellcolor{red!50} 0m40.360s & \cellcolor{red!50} 1m9.523s \\  
                 & bulln & 20 & 61 & \cellcolor{red!50} 15m23.190s & \cellcolor{red!50} 1m19.553s & \cellcolor{red!50} 8m7.918s \\  
                 & camel & 19 & 59 & \cellcolor{red!50} 16m29.513s & \cellcolor{red!50} 3m38.506s & \cellcolor{red!50} 14m34.180s \\  
                 & exxm & 15 & 47 & \cellcolor{red!50} 15m52.984s & \cellcolor{red!50} 3m46.605s & \cellcolor{red!50} 3m43.372s \\  
                 & lion & 18 & 55 & \cellcolor{red!50} 15m37.603s & \cellcolor{red!50} 1m31.160s & \cellcolor{red!50} 4m26.537s \\  
                 & tiger & 17 & 51 & \cellcolor{red!50} 15m50.498s & \cellcolor{red!50} 0m37.245s & \cellcolor{red!50} 1m54.818s \\  
                \hline
            \end{tabular}
            \label{tab: behave against oracle}
            \vspace{-5pt}
            
        \end{table}

\subsubsection{Structural Solution}
To evaluate \struct, we generate locked versions of ISCAS’89~\cite{iscas85} and ITC’99~\cite{itc99}. The encryption is done in .bench format with our Python implementation of \struct~and tested against NEOS~\cite{neos} and RANE~\cite{RANE} attacks. None of the benchmarks run provide the correct keys as shown in Table~\ref{tab: struct against oracle guided}.

        \begin{table}[!t]
          \caption{\struct~security against logic attacks}
          \centering
          \setlength{\tabcolsep}{2pt} % Reduce the space between columns
          \tiny        
          \begin{tabular}{|c|l|c|c|c|c|c|c|}
              \hline
               \multicolumn{4}{|c|}{\textbf{Benchmark and Locking Information}} & \multicolumn{3}{|c|}{\textbf{NEOS~\cite{neos}}} & \textbf{RANE~\cite{RANE}} \\
              \hline
               \textbf{} & \textbf{Circuit} & \textbf{\# keys ($k$) } & \textbf{Key Size ($k_i$)} & \textbf{BBO} & \textbf{INT} & \textbf{KC2} & \textbf{RANE} \\
              \hline
              \multirow{12}{*}{\rotatebox[origin=c]{90}{\textbf{ISCAS'89~\cite{iscas85}}}}
                & s1196 & 4 & 14 & \cellcolor{red!50} 1m20.096s & \cellcolor{red!50} 0m0.694s & \cellcolor{red!50} 0m0.753s & \cellcolor{red!50}  0m1.667s \\  
                & s13207 & 8 & 31 & \cellcolor{red!50} 15m5.270s & \cellcolor{red!50} 0m15.520s & \cellcolor{red!50} 0m19.852s & \cellcolor{red!50}  0m35.909s \\  
                & s1488 & 2 & 8 & \cellcolor{red!50} 1m37.663s & \cellcolor{red!50} 0m0.672s & \cellcolor{red!50} 0m0.723s & \cellcolor{red!50}  0m1.224s \\  
                & s15850 & 4 & 14 & \cellcolor{red!50} 15m9.218s & \cellcolor{red!50} 0m12.460s & \cellcolor{red!50} 0m14.394s & \cellcolor{red!50}  0m20.279s \\  
                & s298 & 2 & 3 & \cellcolor{red!90} 0m0.043s & \cellcolor{red!50} 0m0.474s & \cellcolor{red!50} 0m0.474s & \cellcolor{red!50}  0m0.798s \\  
                & s349 & 4 & 9 & \cellcolor{red!50} 7m21.210s & \cellcolor{red!50} 0m0.672s & \cellcolor{red!50} 0m0.695s & \cellcolor{red!50}  0m0.964s \\  
                & s35932 & 8 & 35 & \cellcolor{red!50} 15m5.694s & \cellcolor{red!50} 3m43.671s & \cellcolor{red!50} 4m9.463s & \cellcolor{red!50}  3m31.814s \\  
                & s510 & 8 & 19 & \cellcolor{red!50} 0m35.772s & \cellcolor{red!50} 0m0.539s & \cellcolor{red!50} 0m0.540s & \cellcolor{red!50}  0m0.942s \\  
                & s5378 & 8 & 35 & \cellcolor{red!50} 7m50.965s & \cellcolor{red!50} 0m1.287s & \cellcolor{red!50} 0m1.486s & \cellcolor{red!50}  0m3.591s \\  
                & s641 & 8 & 35 & \cellcolor{red!50} 0m58.063s & \cellcolor{red!50} 0m0.804s & \cellcolor{red!50} 0m1.191s & \cellcolor{red!50}  0m1.326s \\  
                & s713 & 8 & 35 & \cellcolor{red!50} 0m56.985s & \cellcolor{red!50} 0m0.624s & \cellcolor{red!50} 0m0.659s & \cellcolor{red!50}  0m1.234s \\  
                & s832 & 8 & 18 & \cellcolor{red!50} 0m49.561s & \cellcolor{red!50} 0m0.563s & \cellcolor{red!50} 0m0.603s & \cellcolor{red!50}  0m1.080s \\  
                & s9234 & 8 & 19 & \cellcolor{red!50} 15m4.725s & \cellcolor{red!50}  6h44m50s & \cellcolor{red!50}  7h56m45s & \cellcolor{red!50}50m 6.04s \\
                & s953 & 4 & 15 & \cellcolor{red!50} 0m52.608s & \cellcolor{red!50} 0m0.826s & \cellcolor{red!50} 0m0.127s & \cellcolor{red!90} 2h6m4.59s \\ 
              \hline
                \multirow{20}{*}{\rotatebox[origin=c]{90}{\textbf{ITC'99~\cite{itc99}}}}
               & b01 & 2 & 2 & \cellcolor{red!90} 0m0.296s & \cellcolor{red!50}  0m1.023s & \cellcolor{red!50} 0m0.882s & \cellcolor{red!50} 9m6.02s\\ 
               & b02 & 2 & 2 & \cellcolor{red!90} 0m0.143s & \cellcolor{red!50}  0m0.487s & \cellcolor{red!50} 0m0.653s & \cellcolor{red!50} 10m39.54s\\ 
               & b03 & 2 & 4 & \cellcolor{red!50} 15m0.528s & \cellcolor{red!50}  0m0.473s & \cellcolor{red!50} 0m0.653s & \cellcolor{red!50} 13m6.39s\\ 
               & b04 & 4 & 11 & \cellcolor{red!50} 0m52.426s & \cellcolor{red!50}  0m0.820s & \cellcolor{red!90} 0m0.194s & \cellcolor{red!50} 4h5m53.21s\\ 
               & b05 & 2 & 2 & \cellcolor{red!90} 0m0.153s & \cellcolor{red!90}  0m0.097s & \cellcolor{red!90} 0m0.089s & \cellcolor{red!20} 0m0.415s \\ 
               & b06 & 2 & 1 & \cellcolor{red!90} 0m0.151s & \cellcolor{red!50}  0m0.402s & \cellcolor{red!90} 0m0.165s & \cellcolor{red!50} 0m0.441s \\ 
               & b07 & 2 & 2 & \cellcolor{red!90} 0m0.163s & \cellcolor{red!50}  0m0.739s & \cellcolor{red!50} 0m0.863s & \cellcolor{red!50} 0m0.544s \\ 
               & b08 & 4 & 9 & \cellcolor{red!50} 0m14.811s & \cellcolor{red!50}  0m0.600s & \cellcolor{red!90} 0m0.186s & \cellcolor{red!50} 14m34.59s \\ 
               & b09 & 2 & 1 & \cellcolor{red!90} 0m0.250s & \cellcolor{red!50}  0m0.658s & \cellcolor{red!50} 0m0.698s & \cellcolor{red!50} 0m0.560s \\ 
               & b10 & 4 & 11 & \cellcolor{red!50} 0m16.103s & \cellcolor{red!50}  0m0.719s & \cellcolor{red!90} 0m0.206s & \cellcolor{red!50} 16m48.31s \\ 
               & b11 & 2 & 7 & \cellcolor{red!50} 1m36.385s & \cellcolor{red!50}  0m1.699s & \cellcolor{red!90} 0m0.256s & \cellcolor{red!50} 23m17.52s \\ 
               & b12 & 2 & 5 & \cellcolor{red!50} 16m20.762s & \cellcolor{red!50}  1m24.733s & \cellcolor{red!90} 0m0.261s & \cellcolor{red!50} 1h27m50.43s \\ 
               & b14 & 8 & 32 & \cellcolor{red!50} 15m3.473s & \cellcolor{red!50}  1m55.654s & \cellcolor{red!90} 0m1.083s & \cellcolor{red!50} 19m39.38s \\ 
               & b15 & 16 & 36 & \cellcolor{red!50} 14m3.219s & \cellcolor{red!90}  20m0.006s & \cellcolor{red!90} 0m4.006s & \cellcolor{red!50} 40m34.59s \\ 
               & b17 & 16 & 37 & \cellcolor{red!50} 17m1.496s & \cellcolor{red!20}  20m0.008s & \cellcolor{red!20} 20m0.011s & \cellcolor{yellow!100} 20h0m0.35s \\ 
               & b18 & 16 & 37 & \cellcolor{red!90} 0m0.320s & \cellcolor{red!90}  0m0.258s & \cellcolor{red!50} 0m0.252s & \cellcolor{red!50} 1h59m18.06s \\ 
               & b19 & 8 & 24 & \cellcolor{red!90} 0m0.538s & \cellcolor{red!90}  0m0.574s & \cellcolor{red!50} 0m0.752s & \cellcolor{red!50} 18h6m12.74s \\ 
               & b20 & 8 & 32 & \cellcolor{red!50} 15m6.045s & \cellcolor{red!50}  6m20.914s & \cellcolor{red!20} 0m4.988s & \cellcolor{red!50} 0m57.046s \\ 
               & b21 & 8 & 32 & \cellcolor{red!50} 15m11.598s & \cellcolor{red!50}  6m49.946s & \cellcolor{red!90} 0m5.389s & \cellcolor{red!50} 0m59.616s \\ 
               & b22 & 8 & 32 & \cellcolor{red!50} 15m24.620s & \cellcolor{red!90}  20m0.005s & \cellcolor{red!90} 2m41.473s & \cellcolor{red!50} 2m24.392s \\ 
              \hline
          \end{tabular}
          \label{tab: struct against oracle guided}
        \vspace{-10pt}
    \end{table}

       % This image should be placed here so that it appears in the correct page
        \begin{figure*}[]
            \centering
            \subfloat[Power (W)]{\includegraphics[width=\columnwidth]{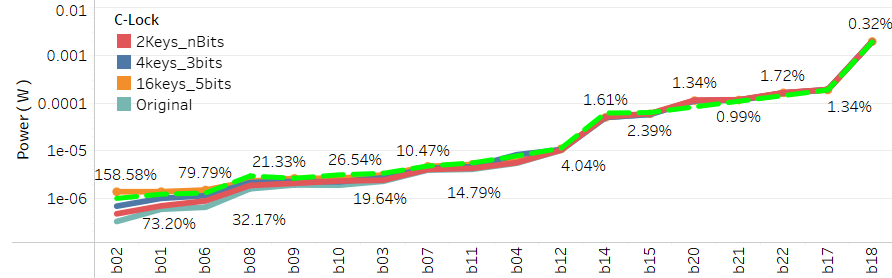}\label{fig:power}} \hspace{0.5em}
            \subfloat[Area ($\mu m^2$)]{\includegraphics[width=\columnwidth]{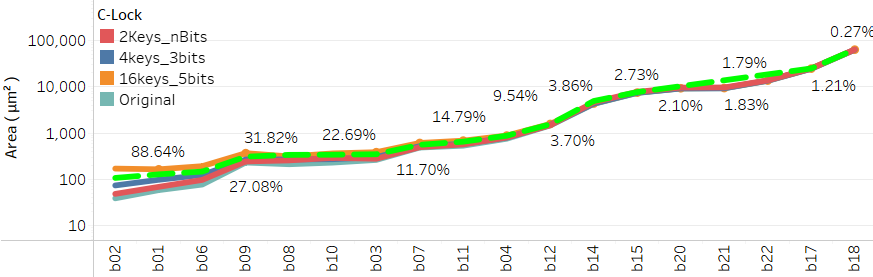}\label{fig:area}} \\ 
            \vspace{-2pt}
            \subfloat[Cell Count]{\includegraphics[width=\columnwidth]{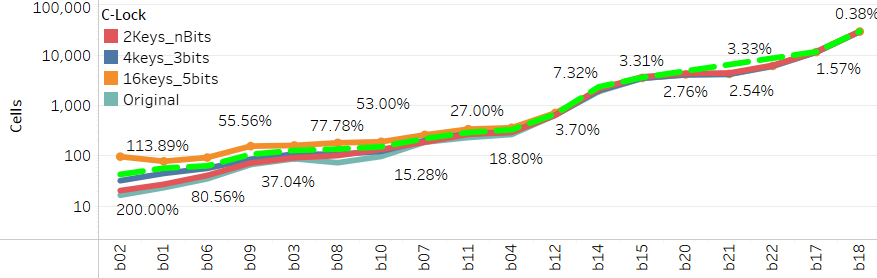}\label{fig:cells}} \hspace{0.5em}
            \subfloat[Number of IOs]{\includegraphics[width=\columnwidth]{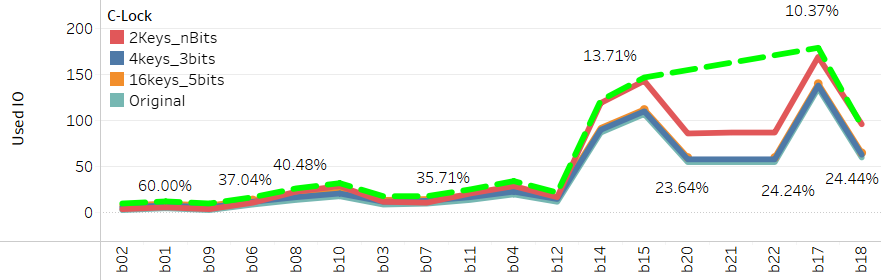}\label{fig:io}} 
            \caption{Overhead comparison of \struct~with DK-Lock~\cite{DKLock}}
            \label{fig:overhead_metrics}
        \end{figure*}

\begin{table}[!b]
\caption{\struct~security against removal attacks}
\centering
\scriptsize
\begin{tabular}{|c|c|c|c|c|c|}
\hline
\multicolumn{1}{|c|}{} & \textbf{DANA \cite{dana} } & \multicolumn{3}{|c|}{\textbf{FALL \cite{fall}}} \\ \hline
\textbf{Circuit} & \textbf{NMI Score} & \textbf{Candidates} & \textbf{Keys} & \textbf{CPU Time (s)} \\ \hline
b01 & 0.00 & 0 & 0 & 0.1 \\ \hline
b02 & 0.00 & 0 & 0 & 0.1 \\ \hline
b03 & 0.00 & 0 & 0 & 0.1 \\ \hline
b04 & 0.00 & 0 & 0 & 0.1 \\ \hline
b05 & 0.00 & 0 & 0 & 0.2 \\ \hline
b06 & 0.00 & 0 & 0 & 0.1 \\ \hline
b07 & 0.74 & 0 & 0 & 0.1 \\ \hline
b08 & 0.99 & 0 & 0 & 0.1 \\ \hline
b09 & 0.43 & 0 & 0 & 0.1 \\ \hline
b10 & 0.00 & 0 & 0 & 0.1 \\ \hline
b11 & 0.76 & 0 & 0 & 0.1 \\ \hline
b12 & 0.99 & 0 & 0 & 0.1 \\ \hline
b14 & 0.60 & 0 & 0 & 10.9 \\ \hline
b15 & 0.89 & 0 & 0 & 16.8 \\ \hline
b17 & 0.93 & 0 & 0 & 97.3 \\ \hline
b18 & 0.93 & 0 & 0 & 1663.6 \\ \hline
b19 & 0.50 & 0 & 0 & 3423.4 \\ \hline
b20 & 0.56 & 0 & 0 & 23.4 \\ \hline
b21 & 0.44 & 0 & 0 & 23.7 \\ \hline
b22 & 0.39 & 0 & 0 & 46.4 \\ \hline
\end{tabular}
\label{tab:removal-attacks}
\end{table}

\subsection{Removal Attacks Evaluation}
\label{Sec:Removal-Attacks-Eval}
As mentioned before, \behave~does not add much benefit to security against removal attacks; however, \struct~allows the circuit to resist against them.

\subsubsection{Dataflow Attack} 
To execute DANA \cite{dana}, we synthesize the ITC’99~\cite{itc99} benchmarks using Xilinx Vivado.  After preparing the netlists, we apply the DANA script to analyze the dataflow and generate the resulting register clusters. DANA does not provide a simple pass/fail output. Instead, it produces clusters that represent potential high-level registers within the circuit. These clusters are then evaluated using the Normalized Mutual Information (NMI) metric, which measures how closely DANA’s output matches the ground truth or, in our case, the original circuit. An NMI value of ``0'' means the tool fails to identify the correct register groupings, while an NMI value of ``1'' means that DANA’s output perfectly matches the reference design. In the original study, DANA was able to get very high NMI scores in the range of 0.87 to 0.99 and an average of 0.95 when compared against the ground truth. When we run DANA against locked benchmarks with \struct, as shown in Table~\ref{tab:removal-attacks}, it is clear that the NMI scores accuracy drops significantly to a wide range of 0.00 to 0.99 and an average of 0.41. These results demonstrate that \struct~changes the dataflow in most of the locked benchmarks compared to the original benchmarks and thus \struct~is able to increase resiliency against dataflow attacks and deteriorate reverse engineering in sequential benchmarks.

\subsubsection{Functional Analysis Attack}
FALL \cite{fall} is designed to work with circuits in .bench format, which we use to test against our locked circuits. In the original study, FALL is reported to be successful against 65 out of 80 locked circuits (81\% success rate). When we run FALL against locked circuits with \struct~on ITC’99 benchmarks \cite{itc99}, the FALL attack fails to find any key (0\% success rate) as shown in Table~\ref{tab:removal-attacks}. This result demonstrates that our \struct~defense is resilient against this type of attack as well.

\subsection{Overhead Analysis}
Now, we look at how much overhead \struct~adds to circuits and compare it with state-of-the-art multi-key logic locking method DK-Lock~\cite{DKLock}. It is worth noting that DK-Lock is not fully secure since it is vulnerable against unrolling attacks such as \cite{Rezaei:BreakUnroll} while as shown in Sections \ref{Sec:Logic-Attacks-Eval} and \ref{Sec:Removal-Attacks-Eval}, \struct~is secure against all existing attack surfaces. 

For the overhead comparison, we focus on four key aspects: power usage, circuit area, number of cells, and number of I/O ports. To evaluate the overhead of \struct, we use circuits from the ITC 99~\cite{itc99} benchmark set. We convert .bench files to Verilog using the ABC tool~\cite{abc}. Then, we use Cadence Genus with a 45nm process to synthesize and get the overhead values. We test the following three configurations:
    \begin{itemize}
        \item \textbf{Test Run 1:} $2$ keys, $n$ bits each ($k=2, k_i=n$)
        \item \textbf{Test Run 2:} 4 keys, 3 bits each 
        \item \textbf{Test Run 3:} 16 keys, 5 bits each 
    \end{itemize}

For DK-Lock, we use two setups: one with 10-bit keys and another where the key size changes linearly based on the inputs to the circuit (i.e., $n=k$). In Figure~\ref{fig:overhead_metrics}, the green dashed line shows the average of these DK-Lock setups. Looking at the figure~\ref{fig:overhead_metrics}, we can see that as circuits get larger, the extra power, area, and cells needed for \struct{} get smaller. This means \struct~scales well for large circuits. While the smallest circuit might use about 100\% more power, for the largest ones, it is less than 1\%. In addition, for large circuits (\textit{b14}-\textit{b22}), both \struct~and DK-Lock do not add much overhead. But for smaller and medium-sized circuits (\textit{b01}-\textit{b11}), our \textbf{Test Run 1} and \textbf{Test Run 2} do a better job than DK-Lock. For example, for the \textit{b06} benchmark, \struct~uses about 30\% less power, area, and cells compared to DK-Lock. It is worth noting that the DK-Lock data does not include the \textit{b20}, \textit{b21}, and \textit{b22} benchmarks, and this is why, in the graphs, there is a line jump. 
        
\section{Conclusion} 
\label{Sec:Conclusion}
In this paper, we introduced \algName, a novel time-based multi-key logic locking family with two variants: \behave~for RTL-level behavioral locking and \struct~for netlist-level structural locking. We demonstrated the resilience of the \algName~family against state-of-the-art oracle-guided SAT attacks incorporated in NEOS \cite{neos} and RANE \cite{RANE} across a wide range of benchmarks. We showed that \struct~improves structural integrity and is resistant to DANA \cite{dana} and FALL \cite{fall} attacks. In addition, we showed that \struct~adds minimal overhead, particularly for large circuits.

Overall, the \struct~effectiveness against both oracle-guided and removal attacks, coupled with its low overhead, makes it a promising practical solution for protecting hardware IPs in the semiconductor supply chain. For future works, multi-key solutions can be explored to address other hardware security problems, such as hardware Trojan detection and mitigation.

\section*{Acknowledgment}
This material is based upon work supported by the National Science Foundation under Award No. 2245247.

\end{document}